\PassOptionsToPackage{table,xcdraw,dvipsnames}{xcolor}
\documentclass[11pt,a4paper]{article}
\usepackage[hyperref]{eacl2021}
\usepackage{times}
\usepackage{latexsym}

\usepackage{microtype}

\usepackage{booktabs}
\usepackage{multirow}
\usepackage{arydshln}
\usepackage{makecell}
\usepackage{graphicx}
\usepackage{amsmath}
\usepackage[normalem]{ulem}

\newcommand{\set}[1]{\mathcal{#1}}

\newcommand{\mantis}{\texttt{MANTiS}}
\newcommand{\msdialog}{\texttt{MSDialog}}
\newcommand{\ubuntu}{\texttt{UDC\textsubscript{DSTC8}}}

\newcommand{\nsbm}{\texttt{NS\textsubscript{BM25}}}
\newcommand{\nsrandom}{\texttt{NS\textsubscript{random}}}
\newcommand{\nssentencebert}{\texttt{NS\textsubscript{sentenceBERT}}}

\newcommand{\crossdataset}{cross-domain}
\newcommand{\crossns}{cross-NS}
\newcommand{\deterministicbert}{\texttt{BERT}}
\newcommand{\rabertdropout}{\texttt{RA-BERT$^{D}$}}
\newcommand{\rabertensemble}{\texttt{RA-BERT$^{E}$}}
\newcommand{\mcdropout}{\texttt{S-BERT$^{D}$}}
\newcommand{\deepensemble}{\texttt{S-BERT$^{E}$}}
\newcommand{\prediction}{$E[R^{D}]$}
\newcommand{\predictionanddropout}{+$var[R^{D}]$}
\newcommand{\predictionandenesmble}{+$var[R^{E}]$}

\newcommand{\cOne}{\textbf{[C1]}}
\newcommand{\cTwo}{\textbf{[C2]}}
\newcommand{\bertRankers}{BERT-based rankers}
\newcommand{\bertRanker}{BERT-based ranker}
\newcommand{\cls}{\texttt{[CLS]}}
\newcommand{\usep}{[U]}
\newcommand{\tsep}{[T]}
\newcommand{\predictivedistribution}{predictive distribution}

\newcommand{\mcd}{\texttt{Dropout}}
\newcommand{\de}{\texttt{Ensemble}}
\newcommand{\ltr}{L2R}

\aclfinalcopy 

\title{On the Calibration and Uncertainty of Neural Learning to Rank Models}

\author{Gustavo Penha \\
  Delft University of Technology \\
  Delft \\
  Netherlands \\
  \texttt{g.penha-1@tudelft.nl} \\\And
  Claudia Hauff \\
  Delft University of Technology \\
  Delft \\
  Netherlands \\
  \texttt{c.hauff@tudelft.nl} \\}

\date{}

\begin{document}
\maketitle

\begin{abstract}

According to the Probability Ranking Principle (PRP), ranking documents in decreasing order of their probability of relevance leads to an optimal document ranking for ad-hoc retrieval. The PRP holds when two conditions are met: \cOne{}~the models are \emph{well calibrated}, and, \cTwo{}~the probabilities of relevance are reported \emph{with certainty}.
We know however that deep neural networks (DNNs) are often not well calibrated and have several sources of uncertainty, and thus \cOne{} and \cTwo{} might not be satisfied by neural rankers. 
Given the success of neural Learning to Rank (\ltr{}) approaches---and here, especially BERT-based approaches---we first analyze under which circumstances deterministic, i.e. outputs point estimates, neural rankers are calibrated. Then, motivated by our findings we use two techniques to model the uncertainty of neural rankers leading to the proposed \emph{stochastic rankers}, which output a \predictivedistribution{} of relevance as opposed to point estimates.
Our experimental results on the ad-hoc retrieval task of conversation response ranking\footnote{The source code and data are available at \url{https://github.com/Guzpenha/transformer_rankers/tree/uncertainty_estimation}.} reveal that (i) \bertRankers{} are not robustly calibrated and that stochastic \bertRankers{} yield better calibration; and (ii) uncertainty estimation is beneficial for both risk-aware neural ranking, i.e. taking into account the uncertainty when ranking documents, and for predicting unanswerable conversational contexts.

\end{abstract}

\section{Introduction}

\definecolor{magenta}{HTML}{FF00FF}
\definecolor{bluebar}{HTML}{0000ff}
\definecolor{grey}{HTML}{666666}
\definecolor{orange}{HTML}{ff9900}

\begin{figure}[]
    \centering
    \includegraphics[width=.48\textwidth]{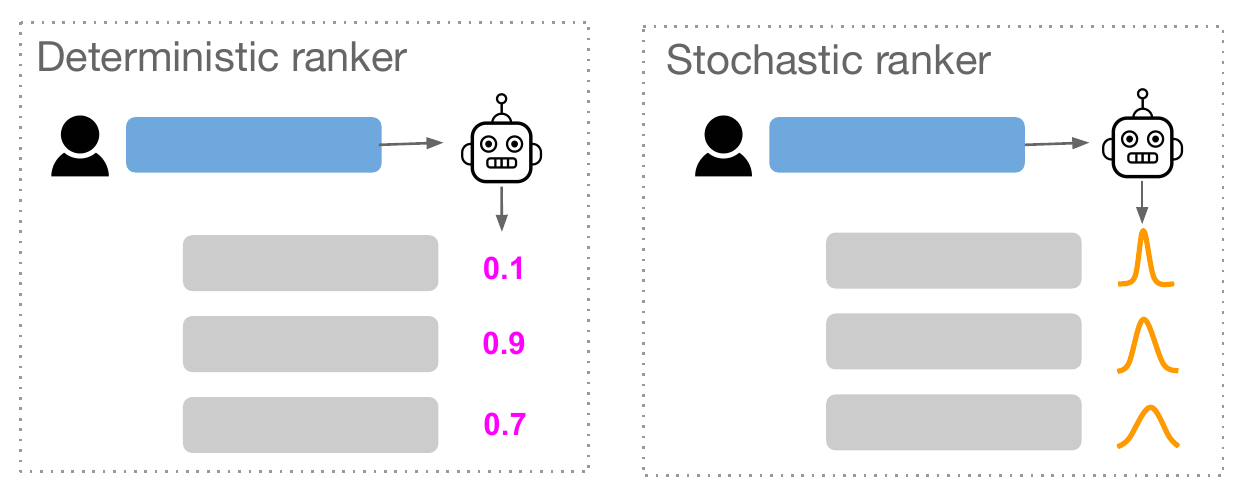}
    \caption{While deterministic neural rankers output a point estimate probability ({\color{magenta}magenta} values) of relevance for a combination of query ({\color{bluebar}blue} bars) and document ({\color{grey}grey} bars), stochastic neural rankers output a \predictivedistribution{} ({\color{orange}orange} curves). The dispersion of the \predictivedistribution{} provides an estimation of the model uncertainty.}
    \label{fig:stochastic_vs_deterministic}    
\end{figure}

According to the Probability Ranking Principle (PRP)~\cite{robertson1977probability}, ranking documents in decreasing order of their probability of relevance leads to an optimal document ranking for ad-hoc retrieval\footnote{Standard retrieval task where the user specifies his information need through a query which initiates a search by the system for documents that are likely relevant~\cite{baeza1999modern}.}. \citet{gordon1991utility} discussed that for the PRP to hold, ranking models must at least meet the following conditions: \cOne{} assign well calibrated probabilities of relevance, i.e. if we gather all documents for which the model predicts relevance with a probability of e.g. 30\%, the amount of relevant documents should be 30\%, and \cTwo{} report certain predictions, i.e. only point estimates such as e.g. 80\% probability of relevance.

DNNs have been shown to outperform classic Information Retrieval (IR) ranking models over the past few years in setups where considerable training data is available. It has been shown that DNNs are not well calibrated in the context of computer vision~\cite{guo2017calibration}. If the same is true for neural \ltr{} models for IR, e.g. transformer-based models for ranking~\cite{nogueira2019passage}, \cOne{} is not met. Additionally, there are a number of sources of uncertainty in the training process of neural networks~\cite{gal2016uncertainty} that make it unreasonable to assume that neural ranking models fulfill \cTwo{}: \emph{parameter uncertainty} (different combinations of weights that explain the data equally well), \emph{structural uncertainty} (which neural architecture to use for neural ranking), and \emph{aleatoric uncertainty} (noisy data). Given these sources of uncertainty, using point estimate predictions and ranking according to the PRP might not achieve the optimal ranking for retrieval. While the effectiveness benefits of risk-aware models~\cite{wang2009mean,wang2009portfolio} which take into account the risk\footnote{In this paper we use risk and uncertainty interchangeably.} have been shown for non-neural IR approaches, this has not yet been explored for neural \ltr{} models.

In this paper we first analyze the calibration of neural rankers, specifically \bertRankers{} for IR tasks to evaluate how calibrated they are. Then, to model the uncertainty of \bertRankers{}, we propose \emph{stochastic} neural ranking models (see Figure~\ref{fig:stochastic_vs_deterministic}), by applying different techniques to model uncertainty of DNNs, namely MC Dropout~\cite{gal2016dropout} and Deep Ensembles~\cite{lakshminarayanan2017simple} which are agnostic to the particular DNN.

In our experiments, we test models under \emph{distributional shift}, i.e. the test data distribution is different from the training data also referred to as out-of-distribution (OOD) examples~\cite{lee2018simple}. In real-world settings, there are often inputs that are shifted due to factors such as non-stationarity and sample bias. Additionally, this experimental setup provides a way of measuring whether the DNN \textit{"know what it knows"}~\cite{ovadia2019can}, e.g. output high uncertainty for OOD examples. 

We find that \bertRankers{} are not robustly calibrated. Stochastic \bertRankers{} have 14\% less calibration error on average than \bertRankers{}. Uncertainty estimation from stochastic \bertRankers{} is advantageous for downstream applications as shown by our experiments for risk-aware neural ranking (2\% more effective on average relative to a model without risk-awareness) and for predicting unanswerable conversational contexts (improves classification by 33\% on average of all conditions).

\section{Related Work}

\subsubsection*{Calibration and Uncertainty in IR}
Even though to optimally rank documents according to the PRP~\cite{robertson1977probability} requires the model to be calibrated~\cite{gordon1991utility} (\cOne{}), the calibration of ranking models has received little attention in IR. In contrast, in the machine learning community there have been a number of studies about calibration~\cite{ovadia2019can,maddox2019simple}, due to the larger decision making pipelines DNNs are often part of and their importance for model interpretability~\cite{thiagarajan2020calibrating}. For instance, in the automated medical domain it is important to provide a calibrated confidence measure besides the prediction of a disease diagnosis to provide clinicians with sufficient information~\cite{jiang2012calibrating}.~\citet{guo2017calibration} has shown that DNNs are not well calibrated in the context of computer vision, motivating our study of the calibration of neural \ltr{} models.

The second condition (\cTwo{}) for optimal retrieval when ranking according to the PRP~\cite{gordon1991utility} is that models report predictions with certainty. While the (un)certainty has not been studied in neural \ltr{} models, there are classic approaches in IR that model the uncertainty. Such approaches have been mostly inspired by economics theory, treating variance as a measure of uncertainty~\cite{varian1999economics}. Following such ideas, non-neural ranking models that take uncertainty into account (i.e. risk-aware models), and thus do not follow the PRP~\cite{robertson1977probability}, have been proposed~\cite{zhu2009risky,wang2009portfolio}, showing significant effectiveness improvements compared to the models that do not model uncertainty. Uncertainty estimation is a difficult task that has other applications in IR besides improving the ranking effectiveness: it can be employed to decide between asking clarifying questions and providing a potential answer in conversational search~\cite{aliannejadi2019asking}; to perform dynamic query reformulation~\cite{lin2020query} for queries where the intent is uncertain; and to predict questions with no correct answers~\cite{feng-etal-2020-none}. 


\subsubsection*{Bayesian Neural Networks}
Unlike standard algorithms to train neural networks, e.g. SGD, that fit point estimate weights given the observed data, Bayesian Neural Networks (BNNs) infer a distribution over the weights given the observed data. \citet{Denker1987LargeAL} contains one of the earliest mentions of choosing probability over weights of a model. An advantage of the Bayesian treatment of neural networks~\cite{mackay1992practical,neal2012bayesian,blundell2015weight} is that they are better at representing existing uncertainties in the training procedure. One limitation of BNNs is that they are computationally expensive compared to DNNs. This has lead to the development of techniques that scale well, and do not require modifications of the neural net architecture and training procedure. \citet{gal2016dropout} proposed a way to approximate Bayesian inference by relying on dropout~\cite{srivastava2014dropout}. While dropout is a regularization technique that ignores units with probability $p$ during every training iteration and is disabled at test time, \mcd{}~\cite{gal2016dropout} employs dropout at both train and test time and generates a \predictivedistribution{} after a number of forward passes. \citet{lakshminarayanan2017simple} proposed an alternative: they employ ensembles of models (\de{}) to obtain a \predictivedistribution{}. \citet{ovadia2019can} showed that \de{} are able to produce well calibrated uncertainty estimates that are robust to dataset shift.

\subsubsection*{Conversational Search}
\emph{Conversational search} is concerned with creating agents that fulfill an information need by means of a \emph{mixed-initiative} conversation through natural language interaction. A popular approach to conversational search is its modeling as an ad-hoc retrieval task: given an ongoing conversation and a large corpus of historic conversations, retrieve the response that is best suited from the corpus (this is also known as conversation response ranking~\cite{wu2017sequential, yang2018response,penha2020curriculum,gu2020speaker,lu2020improving}). This retrieval-based approach does not require task-specific semantics by domain experts~\cite{henderson2019convert}, and it avoids the difficult task of dialogue generation, which often suffers from uninformative, generic responses~\cite{li2016diversity} or responses that are incoherent given the dialogue context~\cite{li2016persona}. One of the challenges of conversational search is identifying unanswerable questions~\cite{feng-etal-2020-none}, which can trigger for instance clarifying questions~\cite{aliannejadi2019asking}. Identifying unanswerable conversational 
contexts is one of the applications we employ uncertainty estimation for. Intuitively, if the system has high uncertainty in the available responses, there may be no correct response available. In this paper we focus on pointwise BERT for ranking, which is a competitive approach for the conversation response ranking task\footnote{\bertRankers{} are currently the top performing models across three conversation response ranking benchmarks: \url{https://bit.ly/34RTJ2r}.}. 
\section{Method}

In this section we introduce the methods used for answering the following research questions: \textbf{RQ1} \textit{How calibrated are deterministic and stochastic \bertRankers{}?} \textbf{RQ2} \textit{Are the uncertainty estimates from stochastic \bertRankers{} useful for risk-aware ranking?} \textbf{RQ3} \textit{Are the uncertainty estimates obtained from stochastic \bertRankers{} useful for identifying unanswerable queries?} We first describe how to measure the calibration of neural rankers (\cOne{}), followed by our approach for modeling and ranking under uncertainty (\cTwo{}), and then we describe how we evaluate their robustness to distributional shift.

\subsection{Measuring Calibration} \label{section:calibration}
To evaluate the \emph{calibration} of neural rankers (\textbf{RQ1}) we resort to the Empirical Calibration Error (ECE)~\cite{naeini2015obtaining}. ECE is an intuitive way of measuring to what extent the confidence scores from neural networks align with the true correctness likelihood. It measures the difference between the observed reliability curve~\cite{degroot1983comparison} and the ideal one. More formally, we sort the predictions of the model, divide them into $c$ buckets $\{B_{0}, ..., B_{c}\}$, and take the weighted average between the average predicted probability of relevance $avg(B_{i})$ and the fraction of relevant documents $\frac{rel(B_{i})}{|B_{i}|}$ in the bucket:
\begin{equation*}
ECE = \sum_{i=0}^{c}  \frac{|B_{i}|}{n} \bigg| avg(B_{i}) - \frac{rel(B_{i})}{|B_{i}|} \bigg| ,
\end{equation*}
where $n$ is the total number of test examples.

\subsection{Modeling Uncertainty}
First we define the ranking problem we focus on, followed by the deterministic \bertRanker{} baseline model (\deterministicbert{}). Having set the foundations, we move to the methods we propose to answer \textbf{RQ2} and \textbf{RQ3}: a stochastic \bertRanker{} to \emph{model uncertainty} (\texttt{S-BERT}) and a risk-aware \bertRanker{} to \emph{take into account uncertainty provided by \texttt{S-BERT} when ranking} (\texttt{RA-BERT}).

\subsubsection{Conversation Response Ranking}
The task of conversation response ranking~\cite{zhang2018modeling,gu2019utterance,tao2019multi,henderson2019convert,penha2020curriculum,yang2020iart} (also known as \emph{next utterance selection}), concerns retrieving the best response given the dialogue context. We choose this specific task due to the large-scale training data available, suitable for the training of neural \ltr{} models. Formally, let $\set{D}=\{(\set{U}_i, \set{R}_i, \set{Y}_i)\}_{i=1}^{N}$ be a data set consisting of $N$ triplets: dialogue context, response candidates and response relevance labels. The dialogue context $\set{U}_i$ is composed of the previous utterances $\{u^1, u^2, ... , u^{\tau}\}$ at the turn $\tau$ of the dialogue. The candidate responses $\set{R}_i = \{r^1, r^2, ..., r^k\}$ are either ground-truth responses or negative sampled candidates, indicated by the relevance labels $\set{Y}_i = \{y^1, y^2, ..., y^k\}$\footnote{Typically, the number of candidates $k \ll K$, where $K$ is the number of available responses and by design the number of ground-truth responses is usually one, the observed response in the conversational data. In our experiments k=10.}. The task is then to learn a ranking function $f(.)$ that is able to generate a ranked list for the set of candidate responses $\set{R}_i$ based on their predicted relevance scores $f(\set{U}_i,r)$. 

\subsubsection{Deterministic \texttt{BERT} Ranker}
We use BERT for learning the function $f(\set{U}_i,r)$, based on the representation learned by the \cls{} token in a pointwise manner. The input for BERT is the concatenation of the context $\set{U}_i$ and the response $r$, separated by SEP tokens. This is the equivalent of early adaptations of BERT for ad-hoc retrieval~\cite{yang2019simple} transported to conversation response ranking. Formally the input sentence to BERT is $concat(\set{U}_i,r) = u^1 \; | \; \usep \; | \; u^2 \; | \;  \tsep \; | \; ... \; | \; u^{\tau} \; | \; [SEP] \; | \; r$, where $|$ indicates the concatenation operation. The utterances from the context $\set{U}_i$ are concatenated with special separator tokens $\usep$ and $\tsep$ indicating end of utterances and turns. The response $r$ is concatenated with the context using BERT's standard sentence separator $[SEP]$. We fine-tune BERT on the target conversational corpus and make predictions as follows: $f(\set{U}_i,r) = \sigma(FFN(BERT_{CLS}(concat(\set{U}_i,r)))),$ where $BERT_{CLS}$ is the pooling operation that extracts the representation of the \cls{} token from the last layer and $FFN$ is a feed-forward network that outputs logits for two classes (relevant and non-relevant). We pass the logits through a softmax transformation $\sigma$ that gives us a probability of relevance. Since $f(\set{U}_i,r)$ outputs a point estimate value of relevance probability, we refer to it as \deterministicbert{}.

\subsubsection{Stochastic \texttt{S-BERT} Ranker}
In order to obtain a \predictivedistribution{}, $R_{r} = \{f(\set{U}_i,r)^{0}, f(\set{U}_i,r)^{1}, ... , f(\set{U}_i,r)^{n}\}$, which allows us to extract uncertainty estimates, we rely on two techniques, namely \de{}~\cite{lakshminarayanan2017simple} and \mcd{}~\cite{gal2016dropout}. Both techniques scale well and do not require modifications on the architecture or training of BERT.

\paragraph{Using Deep Ensembles (\deepensemble{})}
We train $M$ models using different random seeds without changing the training data, each with its own set of parameters $\{\theta_{m}\}_{m=1}^M$ and make predictions with each one of them to generate $M$ predicted values: $R_{r}^{E} = \{f(\set{U}_i,r)^{0}, f(\set{U}_i,r)^{1}, ... , f(\set{U}_i,r)^{M}\}$. The mean of the predicted values is used as the predicted probability of relevance: $\text{\deepensemble{}}(\set{U}_i,r) = E[R_{r}^{E}],$ and the variance $var[R_{r}^{E}]$ gives us a measure of the uncertainty in the prediction.

\paragraph{Using MC Dropout (\mcdropout{})}
We train a single model with parameters $\theta$ and employ dropout at test time and generate stochastic predictions of relevance by conducting $T$ forward passes: $R_{r}^{D} = \{f(\set{U}_i,r)^{0}, f(\set{U}_i,r)^{1}, ... , f(\set{U}_i,r)^{T}\}$. The mean of the predicted values is used as the predicted probability of relevance: $\text{\mcdropout{}}(\set{U}_i,r) = E[R_{r}^{D}],$ and the variance $var[R_{r}^{D}]$ gives us a measure of the uncertainty. 


\subsubsection{Risk-Aware \texttt{RA-BERT} Ranker}
\label{section:riskaware}

Given the \predictivedistribution{} $R_{r}$, obtained either by \de{} or \mcd{}, we use the following function to rank responses with risk-awareness:

\begin{equation*}
    \begin{split}
    \text{\texttt{RA-BERT}}(\set{U}_i,r) = E[R_{r}] - b *  var[R_{r}] \\
    - 2b\sum_{i}^{n-1}cov[R_{r}, R_{r_{i}}],
    \end{split}
\end{equation*}

where $E[R_{r}]$ is the mean of the \predictivedistribution{}, and $b$ is a hyperparameter that controls the aversion or predilection towards risk. Unlike \cite{zuccon2011back}, we are not combining different runs that encompass different model architectures. We instead take a Bayesian interpretation of the process of generating a \predictivedistribution{} from a single model architecture. We refer to the rankers as \rabertdropout{} and \rabertensemble{}, when using \mcdropout{}'s \predictivedistribution{} and \deepensemble{}'s \predictivedistribution{} respectively.

\subsection{Robustness to Distributional Shift}
In order to evaluate whether we can trust the model's calibration and uncertainty estimates, similar to \cite{ovadia2019can} we evaluate how robust the models are to different types of shift in the test data. We do so by training the model using one setting and applying it in a different setting. Specifically for all three research questions we test the models under the following two settings: cross-domain and cross-NS.

\subsubsection{Cross Domain} We train the model using the training set from one domain, i.e. dataset, known as the source domain $\set{D_{S}}$ and evaluate it on the test set of a different domain, known as the target domain $\set{D_{T}}$. This is also known as the problem of domain generalization~\cite{gulrajani2020search}.

\subsubsection{Cross Negative Sampling} Pointwise \ltr{} models are trained pairs of query and relevant document and pairs of query and non relevant documents~\cite{lucchese2017impact}. Selecting the non-relevant documents requires a \emph{negative sampling} (NS) strategy. For the \crossns{} condition, we test models on negative documents that were sampled using a different NS strategy than during training, evaluating the generalization of the models on a shifted distribution of candidate documents. We use three NS strategies. In \textbf{\nsrandom{}} we randomly select a response $r$ from the list of all responses. For \textbf{\texttt{NS\textsubscript{classic}}} we retrieve negative samples using the conversational context $\set{U}_{i}$ as query to a conventional retrieval model and all the responses $r$ as documents. In \textbf{NS\textsubscript{sentenceEmb}} we represent both $\set{U}_{i}$ and all the responses $r$ with a sentence embedding technique and retrieve candidate responses using a similarity measure.
    \definecolor{eceYellow}{HTML}{FFD667}
\definecolor{recallGreen}{HTML}{57BB8A}

\begin{table*}[ht!]
\centering
\scriptsize
\caption{Calibration (ECE, lower is better) and effectiveness (R$_{10}$@1, higher is better) of \deterministicbert{} for conversation response ranking in \crossdataset{}, and \crossns{} conditions. All models were trained using \nsbm. ECE is calculated using a balanced number of relevant and non relevant documents. \uline{Underlined} values indicate no distributional shift ($\set{D_{S}}$ = $\set{D_{T}}$ and train \texttt{NS} = test \texttt{NS}).}
\label{table:effectiveness}
\begin{tabular}{@{}lrrrrrrcrrr@{}}
\toprule
 & \multicolumn{6}{c}{\crossdataset{}} & \multicolumn{4}{c}{\crossns{}} \\ \cmidrule(l){2-7}   \cmidrule(l){8-11} 
Test on $\rightarrow$ & \multicolumn{2}{c}{\mantis{}} & \multicolumn{2}{c}{\msdialog{}} & \multicolumn{2}{c}{\ubuntu{}} & \multicolumn{2}{c}{\nsrandom{}} & \multicolumn{2}{c}{\nssentencebert{}} \\ \cmidrule(r){1-1} \cmidrule(l){2-7}   \cmidrule(l){8-11} 
\makecell{Train on $\downarrow$  \\(\nsbm)} & \multicolumn{1}{c}{R$_{10}$@1} & \multicolumn{1}{c}{ECE} & \multicolumn{1}{c}{R$_{10}$@1} & \multicolumn{1}{c}{ECE} & \multicolumn{1}{c}{R$_{10}$@1} & \multicolumn{1}{c}{ECE} & R$_{10}$@1 & \multicolumn{1}{c}{ECE} & \multicolumn{1}{c}{R$_{10}$@1} & \multicolumn{1}{c}{ECE} \\ \cmidrule(r){1-1} \cmidrule(l){2-7}   \cmidrule(l){8-11} 
\mantis{} & \uline{0.615} & \uline{0.003} & 0.653 & 0.010 & 0.422 & 0.028 & 0.263 & 0.011 & 0.310 & 0.009 \\ 
\msdialog{} & 0.398 & 0.009 & \uline{0.652} & \uline{0.006} & 0.495 & 0.014 & 0.298 & 0.029 & 0.239 & 0.027 \\
\ubuntu{} & 0.349 & 0.016 & 0.306 & 0.023 & \uline{0.834} & \uline{0.002} & 0.318 & 0.050 & 0.182 & 0.045 \\ \bottomrule
\end{tabular}
\end{table*}

    \begin{table*}[ht!]
\centering
\scriptsize
\caption{Relative decreases of ECE (lower is better) of \deepensemble{} and \mcdropout{} over \deterministicbert{}. Superscript $^{\dagger}$ denote significant improvements (95\% confidence interval) using Student's t-tests.}
\label{table:calibration}
\begin{tabular}{@{}lllllllllll@{}}
\toprule
 & \multicolumn{6}{c}{\crossdataset{}} & \multicolumn{4}{c}{\crossns{}} \\ \cmidrule(l){2-7}   \cmidrule(l){8-11} 
Test on $\rightarrow$ & \multicolumn{2}{c}{\mantis{}} & \multicolumn{2}{c}{\msdialog{}} & \multicolumn{2}{c}{\ubuntu{}} & \multicolumn{2}{c}{\nsrandom{}} & \multicolumn{2}{c}{\nssentencebert{}} \\ \cmidrule(r){1-1} \cmidrule(l){2-7}   \cmidrule(l){8-11} 
\makecell{Train on $\downarrow$ \\(\nsbm)} & \multicolumn{1}{l}{\deepensemble{}} & \multicolumn{1}{l}{\mcdropout{}} & \multicolumn{1}{l}{\deepensemble{}} & \multicolumn{1}{l}{\mcdropout{}} & \multicolumn{1}{l}{\deepensemble{}} & \multicolumn{1}{l}{\mcdropout{}} & \multicolumn{1}{l}{\deepensemble{}} & \multicolumn{1}{l}{\mcdropout{}} & \multicolumn{1}{l}{\deepensemble{}} & \multicolumn{1}{l}{\mcdropout{}} \\ \cmidrule(r){1-1} \cmidrule(l){2-7}   \cmidrule(l){8-11} 
\mantis{} & $\textbf{-}$35.13\%$^{\dagger}$ & $\textbf{-}$56.14\%$^{\dagger}$ & $\textbf{-}$03.42\% & $\textbf{-}$26.89\%$^{\dagger}$ & $\textbf{-}$04.94\% & $\textbf{-}$00.83\% & $\textbf{-}$31.35\% & $\textbf{-}$18.65\%$^{\dagger}$ & $\textbf{-}$37.65\%$^{\dagger}$ & $\textbf{-}$02.79\% \\
\msdialog{} & {\tiny +}25.05\% & {\tiny +}08.27\% & $\textbf{-}$43.11\% & $\textbf{-}$11.54\% & {\tiny +}22.77\% & {\tiny +}05.85\% & $\textbf{-}$15.91\% & $\textbf{-}$10.58\% & $\textbf{-}$17.17\% & $\textbf{-}$12.93\% \\
\ubuntu{} & $\textbf{-}$54.95\%$^{\dagger}$  & $\textbf{-}$09.98\%$^{\dagger}$ & $\textbf{-}$25.78\%$^{\dagger}$  & $\textbf{-}$09.15\% & {\tiny +}24.77\% & $\textbf{-}$01.84\% & $\textbf{-}$08.05\% & $\textbf{-}$01.78\% & $\textbf{-}$04.81\% & $\textbf{-}$01.28\% \\ \bottomrule
\end{tabular}
\end{table*}

\section{Experimental Setup}

We consider three large-scale information-seeking conversation datasets\footnote{\msdialog{} is available at~\url{https://ciir.cs.umass.edu/downloads/msdialog/}; \mantis{} is available at~\url{https://guzpenha.github.io/MANtIS/}; \ubuntu{} is available at ~\url{https://github.com/dstc8-track2/NOESIS-II}.} that allow the training of neural ranking models for conversation response ranking: \textbf{\msdialog{}}~\cite{qu2018analyzing} contains 246K context-response pairs, built from 35.5K information seeking conversations from the Microsoft Answer community, a QA forum for several Microsoft products; \textbf{\mantis{}}~\cite{penha2019introducing} contains 1.3 million context-response pairs built from conversations of 14 Stack Exchange sites, such as \textit{askubuntu} and \textit{travel}; \textbf{\ubuntu{}}~\cite{Kummerfeld_2019} contains 184k context-response pairs of disentangled Ubuntu IRC dialogues. 


\subsection{Implementation Details}
We fine-tune BERT~\cite{devlin2019bert} (\textit{bert-base-cased}) for conversation response ranking using the \textit{huggingface-transformers}~\cite{wolf2019huggingface}. We follow recent research in IR that employed fine-tuned BERT for retrieval tasks~\cite{nogueira2019passage,yang2019simple}, including conversation response ranking~\cite{penha2020curriculum,vig2019comparison,whang2019domain}. When training BERT we employ a balanced number of relevant and non-relevant---sampled using BM25~\cite{robertson1994some}---context and response pairs. The sentence embeddings we use for \crossns{} is \texttt{sentenceBERT}~\cite{reimers2019sentence} and we employ dot product calculation from FAISS~\cite{JDH17}. We consider each dataset as a different domain for \crossns{}. We use cross entropy loss and the Adam optimizer~\cite{kingma2014adam} with $lr=5^{-6}$ and $\epsilon = 1^{-8}$, we train with a batch size of $6$ and fine-tune the model for 1 epoch. This baseline \bertRanker{} setup yields comparable effectiveness with SOTA methods\footnote{We obtain 0.834 $R_{10}@1$ on \ubuntu{} with our baseline BERT model, c.f. Table~\ref{table:effectiveness}, while \texttt{SA-BERT}~\cite{gu2020speaker} achieves 0.830. The best performing model of the DSTC8~\cite{kim2019eighth} also employed a fine-tuned BERT}.

\subsection{Evaluation}
To evaluate the \emph{effectiveness} of the neural rankers we resort to a standard evaluation metric in conversation response ranking~\cite{yuan2019multi,gu2020speaker,tao2019multi}: recall at position $K$ with $n$ candidates\footnote{For example $R_{10}@1$ indicates the number of relevant responses found at the first position when the model has to rank 10 candidate responses.}: $R_n@K$. To evaluate the \emph{calibration} of the models, we resort to the Empirical Calibration Error (cf. \S\ref{section:calibration}, using $C=10$). Throughout, we report the test set results for each dataset. To evaluate the \emph{quality of the uncertainty estimation} we rely on two downstream tasks. The first is to improve conversation response ranking itself via Risk-Aware ranking (cf. \S\ref{section:riskaware}). The second, which fits well with conversation response ranking, is to predict unanswerable conversational contexts. Formally the task is to predict whether there is a correct answer in the candidates list $\set{R}$ or not. In our experiments, for half of the instances we remove the relevant response from the list, setting the label as None Of The Above (NOTA). The other half of the data has label 0 indicating that there is a suitable answer in the candidates list, for which we remove one of the negative samples instead. Similar to~\citet{feng-etal-2020-none}, who proposed to use the outputs (logits) of a LSTM-based model in order to predict NOTA, we use the uncertainties as additional features to the classifier for NOTA prediction. The input space with the additional features is fed to a learning algorithm (Random Forest), and we evaluate it with a 5 fold cross-validation procedure using F1-Macro.

\section{Results}

\begin{figure}[]
    \centering
     \includegraphics[width=.4\textwidth]{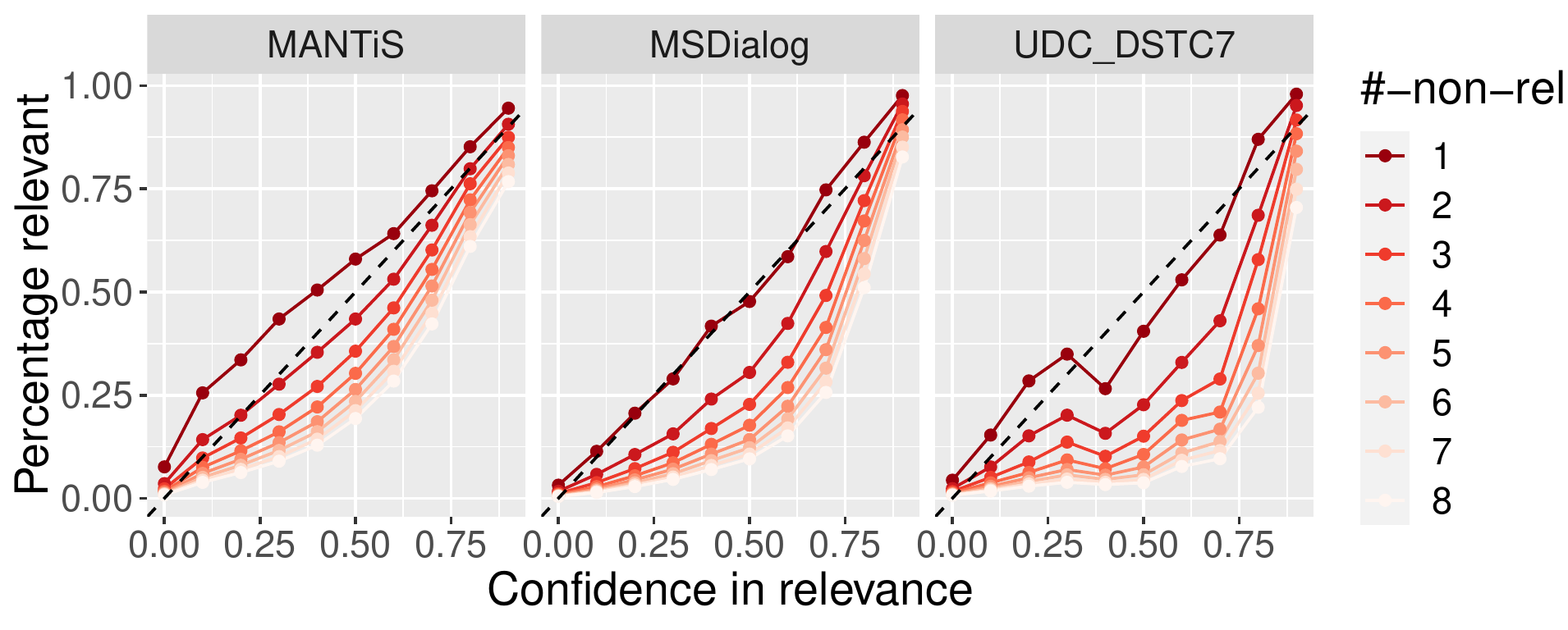}
    \caption{Calibration of \deterministicbert{} trained on a balanced number of relevant and non-relevant documents, and tested on unbalanced data with more non-relevant (\texttt{\#-non-rel}) than relevant (1 per query) documents. A fully calibrated model is represented by the dotted diagonal---for every bucket of confidence in relevance, the percentage of relevant documents found in that bucket is the confidence.}
    \label{fig:calibration_by_n_docs}    
    \vspace{-0.5cm}
\end{figure}

\begin{table*}[ht!]
\centering
\scriptsize
\caption{Relative improvements (higher is better) of $R_{10}@1$ of \rabertensemble{} and \rabertdropout{} over the mean of stochastic BERT predictions (\deepensemble{} and \mcdropout{}). Superscript $^{\dagger}$ denote statistically significant improvements over the \texttt{S-BERT} ranker at 95\% confidence interval using Student's t-tests.}
\label{table:risk-aware-ranking}
\begin{tabular}{@{}lllllllllll@{}}
\toprule
 & \multicolumn{6}{c}{\crossdataset{}} & \multicolumn{4}{c}{\crossns{}} \\ \cmidrule(l){2-7}   \cmidrule(l){8-11} 
Test on $\rightarrow$ & \multicolumn{2}{c}{\mantis{}} & \multicolumn{2}{c}{\msdialog{}} & \multicolumn{2}{c}{\ubuntu{}} & \multicolumn{2}{c}{\nsrandom} & \multicolumn{2}{c}{\nssentencebert} \\ \cmidrule(r){1-1} \cmidrule(l){2-7}   \cmidrule(l){8-11} 
\makecell{Train on $\downarrow$ \\(\nsbm)} & \multicolumn{1}{l}{\tiny{\rabertensemble{}}} & \multicolumn{1}{l}{\tiny{\rabertdropout{}}} & \multicolumn{1}{l}{\tiny{\rabertensemble{}}} & \multicolumn{1}{l}{\tiny{\rabertdropout{}}} & \multicolumn{1}{l}{\tiny{\rabertensemble{}}} & \multicolumn{1}{l}{\tiny{\rabertdropout{}}} & \multicolumn{1}{l}{\tiny{\rabertensemble{}}} & \multicolumn{1}{l}{\tiny{\rabertdropout{}}} & \multicolumn{1}{l}{\tiny{\rabertensemble{}}} & \multicolumn{1}{l}{\tiny{\rabertdropout{}}} \\ \cmidrule(r){1-1} \cmidrule(l){2-7}   \cmidrule(l){8-11} 
\mantis{}   & $\textbf{-}$0.14\% & {\tiny +}0.16\%$^{\dagger}$ & {\tiny +}0.00\%                          & {\tiny +}0.00\%                          & {\tiny +}0.00\% & {\tiny +}0.00\%                          & {\tiny +}4.73\%$^{\dagger}$ & {\tiny +}4.58\%$^{\dagger}$ & {\tiny +}9.68\%$^{\dagger}$  & $\textbf{-}$2.68\% \\
\msdialog{} & $\textbf{-}$2.74\% & {\tiny +}0.39\% & $\textbf{-}$1.05\% & $\textbf{-}$0.66\% & {\tiny +}5.08\%$^{\dagger}$ & $\textbf{-}$0.10\% & $\textbf{-}$7.61\%   & {\tiny +}3.29\% & $\textbf{-}$0.61\% & {\tiny +}0.63\%  \\
\ubuntu{}   & {\tiny +}0.00\%  & {\tiny +}0.00\% & {\tiny +}0.00\%                          & {\tiny +}0.00\%                          & {\tiny +}0.42\% & $\textbf{-}$0.06\%                         & {\tiny +}6.32\%$^{\dagger}$ & {\tiny +}3.83\%$^{\dagger}$ & {\tiny +}16.39\%$^{\dagger}$ & {\tiny +}17.18\%$^{\dagger}$\\ \bottomrule
\end{tabular}
\end{table*}
\begin{figure*}
    \centering
     \includegraphics[width=0.9\textwidth]{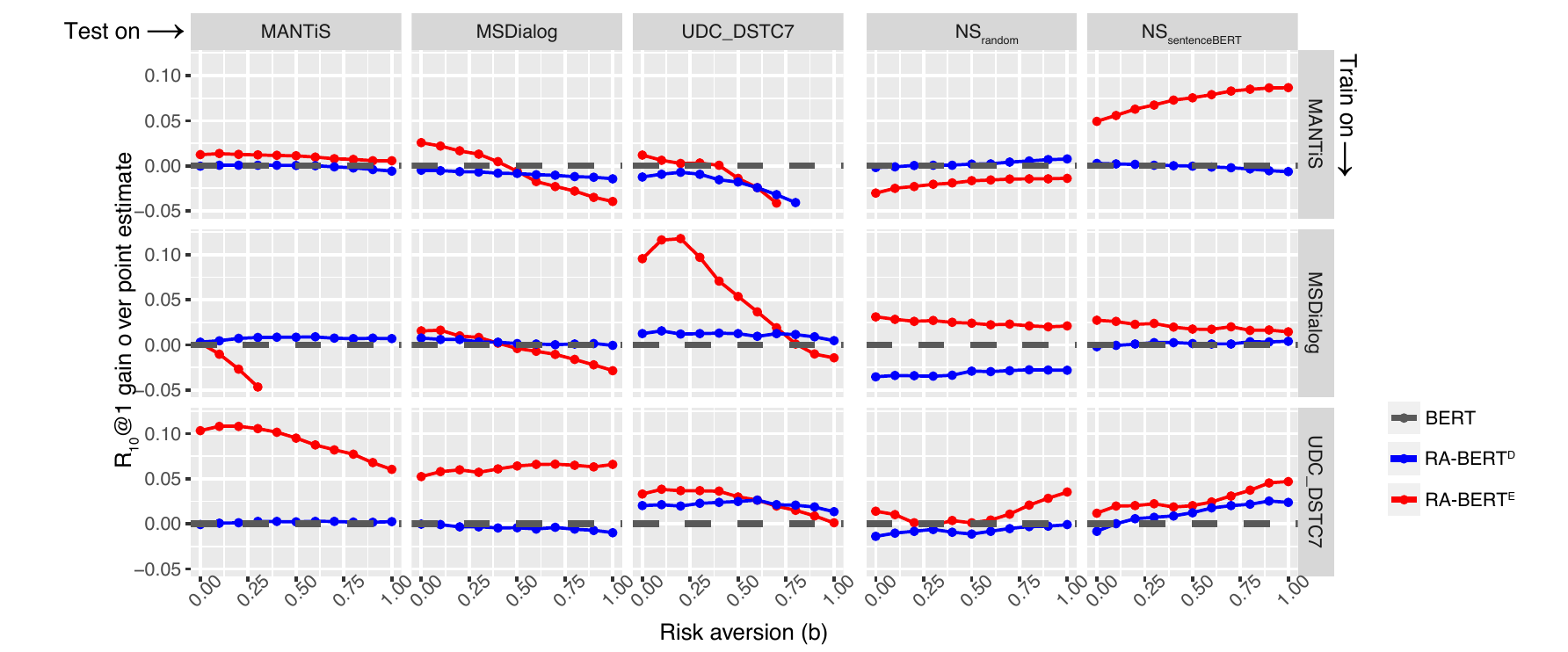}
    \caption{Gains of the Risk-Aware BERT-ranker for different values of risk aversion $b$.}
    \label{fig:risk_aware_by_b}
\end{figure*}

\subsection{Calibration of Neural Rankers (RQ1)}

In order to answer our first research question about the calibration of neural rankers, let us first analyze \deterministicbert{} under standard settings (no distributional shift). Our results show that \deterministicbert{} is both effective and calibrated under no distributional shift conditions. In Table~\ref{table:effectiveness} we see that when the target data (\textit{Test on $\rightarrow$}) is the same as the source data (\textit{Train on $\downarrow$})---indicated by underlined values---we obtain the highest effectiveness (on average 0.70 $R_{10}@1$) and the lowest calibration error (on average 0.036 ECE). When plotting the calibration curves of the model in Figure~\ref{fig:calibration_by_n_docs}, we observe the curves to be almost diagonal (i.e. having near perfect calibration) when there are an equal number of relevant and non-relevant candidates ($\texttt{\#-non-rel} = 1$).

However, when we make the conditions more realistic\footnote{In a production system, the retrieval stage would be executed over all candidate responses. As a consequence, the data is highly unbalanced, i.e. only a few relevant responses among potentially millions of non-relevant responses.} by having multiple non-relevant candidates for each conversational context, we observe in Figure~\ref{fig:calibration_by_n_docs} that the calibration errors start to increase, moving away from the diagonal. Additionally, when we challenge the model in \crossdataset{} and \crossns{} settings, the calibration error increases significantly as evident in Table~\ref{table:effectiveness}. On average, the ECE is \textbf{4.6} times higher for \crossdataset{} and \textbf{7.9} times higher for \crossns{}. Thus \textbf{answering the first part of our first research question about the calibration of deterministic \bertRankers{}, indicating that they do not have robust calibrated predictions}, failing on the scenarios where there is a distributional shift.

In order to answer the remaining part of RQ1, on how calibrated are \emph{stochastic} \bertRankers{}, we consider Table~\ref{table:calibration}. It displays the improvements (relative drop in ECE) over \deterministicbert{} in terms of calibration. We see that \deepensemble{} is on average \textbf{14}\% better (has less calibration error) than \deterministicbert{}, while \mcdropout{} is on average \textbf{10}\% better than \deterministicbert{}, \textbf{answering our first research question: stochastic \bertRankers{} have better calibration than deterministic \bertRanker{}}. We hypothesize that \deepensemble{} lead to less ECE than \mcdropout{} because it better captures the model uncertainty in the training procedure, since it combines different weights that explain equally well the prediction of relevance given the inputs. On the next section we focus on evaluating the effectiveness of such models that are better calibrated and also taking into account uncertainty when ranking.

\begin{table*}[ht!]
\centering
\scriptsize
\caption{Results of the \textit{\crossdataset{}} condition for the NOTA prediction task, using a Random Forest classifier and different input spaces. The F1-Macro and standard deviation over the 5 folds of the cross validation are displayed. Superscript $^{\dagger}$ denote statistically significant improvements over \prediction{} at 95\% confidence interval using Student's t-tests. Bold indicates the most effective approach.}
\label{table:nota_cross_domain}
\begin{tabular}{@{}llllllllll@{}}
\toprule
 & \multicolumn{9}{c}{\crossdataset{}} \\ \cmidrule(l){2-10}
Test on $\rightarrow$ & \multicolumn{3}{c}{\mantis{}} & \multicolumn{3}{c}{\msdialog{}} & \multicolumn{3}{c}{\ubuntu{}} \\
\cmidrule(r){1-1} \cmidrule(l){2-4} \cmidrule(l){5-7} \cmidrule(l){8-10} 
\makecell{Train on $\downarrow$ \\(\nsbm)} & \prediction{} & \predictionandenesmble{} & \predictionanddropout{} & \prediction{} & \predictionandenesmble{} & \predictionanddropout{} & \prediction{} & \predictionandenesmble{} & \predictionanddropout{} \\ \cmidrule(r){1-1} \cmidrule(l){2-4} \cmidrule(l){5-7} \cmidrule(l){8-10} 
\mantis{}   & 0.635 (.02) & 0.686 (.01)$^{\dagger}$ & \textbf{0.792 (.02)}$^{\dagger}$ & 0.669 (.03) & 0.731 (.04)  & \textbf{0.855 (.02)}$^{\dagger}$ & 0.571 (.04) & 0.590 (.08)$^{\dagger}$ & \textbf{0.621 (.04)}$^{\dagger}$  \\
\msdialog{} & 0.561 (.02) & 0.598 (.02)$^{\dagger}$ & \textbf{0.633 (.02)}$^{\dagger}$ & 0.662 (.04) & \textbf{0.702 (.01)}$^{\dagger}$ &  0.699 (.06)$^{\dagger}$ & 0.596 (.04) & 0.566 (.06)$^{\dagger}$ & \textbf{0.655 (.06)}$^{\dagger}$ \\
\ubuntu{}   & 0.527 (.04) & 0.665 (.02)$^{\dagger}$ & \textbf{0.738 (.03)}$^{\dagger}$ & 0.523 (.05) &0.691 (.03)$^{\dagger}$  & \textbf{0.757 (.04)}$^{\dagger}$ & 0.787 (.01) & \textbf{0.829 (.03)}$^{\dagger}$ & 0.807 (.01)$^{\dagger}$ \\ \bottomrule
\end{tabular}
\end{table*}
\begin{table*}[ht!]
\centering
\scriptsize
\caption{Results of the \textit{\crossns{}} condition for the NOTA prediction task.}
\label{table:nota_cross_ns}
\begin{tabular}{@{}lllllll@{}}
\toprule
 & \multicolumn{6}{c}{\crossns{}} \\ \midrule
Test on $\rightarrow$ & \multicolumn{3}{c}{\nsrandom{}} & \multicolumn{3}{c}{\nssentencebert{}} \\ \cmidrule(r){1-1} \cmidrule(l){2-4} \cmidrule(l){5-7} 
\makecell{Train on $\downarrow$ \\(\nsbm)} & \prediction{} & \predictionandenesmble{} & \predictionanddropout{} & \prediction{} & \predictionandenesmble{} & \predictionanddropout{} \\ \cmidrule(r){1-1} \cmidrule(l){2-4} \cmidrule(l){5-7} 
\mantis{}   & 0.557 (.01) & 0.604 (.02)$^{\dagger}$ &  \textbf{0.698 (.02)}$^{\dagger}$ & 0.534 (.03) & 0.587 (.02)$^{\dagger}$ & \textbf{0.647 (.05)}$^{\dagger}$ \\
\msdialog{} & 0.505 (.02) & 0.606 (.02)$^{\dagger}$ & \textbf{0.702 (.05)}$^{\dagger}$ & 0.522 (.03) & 0.611 (.07)$^{\dagger}$ & \textbf{0.653 (.04)}$^{\dagger}$ \\
\ubuntu{}   & 0.565 (.03) & 0.800 (.02)$^{\dagger}$ & \textbf{0.942 (.04)}$^{\dagger}$ & 0.506 (.05) & 0.755 (.05)$^{\dagger}$  & \textbf{0.821 (.05)}$^{\dagger}$\\ \bottomrule
\end{tabular}
\end{table*}

\subsection{Uncertainty Estimates for Risk-Aware Neural Ranking (RQ2)}

In order to evaluate the quality of the uncertainty estimations, we first resort to using them as a measure of the risk through risk-aware neural ranking (\rabertdropout{} and \rabertensemble{}). Figure \ref{fig:risk_aware_by_b} displays the effectiveness in terms of $R_{10}@1$ gains over \deterministicbert{} for the different settings (\crossdataset{} and \crossns{}) when varying the risk aversion $b$. 

We note that when $b=0$, we are using the mean of the \predictivedistribution{} and disregard the risk, which is equivalent to \mcdropout{} and \deepensemble{}. The ensemble based average \deepensemble{} is more effective than the baseline \deterministicbert{} for almost all combinations and \mcdropout{} is equivalent to the baseline. When using $b<0$, we are ranking with risk predilection (the opposite of risk aversion), and in all conditions we found that the effectiveness was significantly worse than when $b=0$ and thus $b<0$ is not displayed in Figure \ref{fig:risk_aware_by_b}.

When increasing the risk aversion ($b>0$), we see that it has different effects depending on the combination of domain and NS. For instance, when training in \msdialog{} and applying on \ubuntu{}, increasing the risk aversion improves effectiveness of \rabertensemble{} until $b$ reaches 0.25 and after that the effectiveness drops, meaning that too much risk aversion is not effective. In order to investigate whether ranking with risk aversion is more effective than using the \predictivedistribution{} mean, we select $b$ based on the best value observed on the validation set. 
Table ~\ref{table:risk-aware-ranking} displays the results of this experiment, showing the improvements of \rabertdropout{} and \rabertensemble{} over \mcdropout{} and \deepensemble{} respectively. The results show that in a few cases (8 out of 30) the best value of $b$ is 0, for which risk-aversion is not the best option in the development set. We obtain effectiveness improvements primarily on the \crossns{} condition (up to 17.2\% improvement of $R_{10}@1$), which is the hardest condition (when the models are most ineffective, c.f. Table~\ref{table:effectiveness}). \textbf{This answers our third research question, indicating that the uncertainties obtained from stochastic neural rankers are useful for risk-aware ranking, specially in the \crossns{} setting where the baseline model is quite ineffective.} \rabertensemble{} is on average 2\% more effective than \deepensemble{}, while \rabertdropout{} is on average 1.7\% more effective than \mcdropout{}.

\subsection{Uncertainty Estimates for NOTA prediction (RQ3)}

Besides using the uncertainty estimation for risk-aware ranking, we also employ it for the NOTA (None of the Above) prediction task. We compare here different input spaces for the NOTA classifier. \prediction{} stands for the input space that only uses the mean of the predictive distribution for the $k$ candidate responses in $\set{R}$ using \mcdropout{}, \predictionandenesmble{} uses both \prediction{} and the uncertainties of \deepensemble{} for the $k$ candidates and \predictionanddropout{} uses both the scores \prediction{} and the uncertainties of \mcdropout{}. Our results show that the uncertainties from \mcdropout{} and of \deepensemble{} significantly improve the F1 for NOTA prediction for both \crossdataset{} (Table~\ref{table:nota_cross_domain}, improvement of 24\% on average when using \mcdropout{}) and \crossns{} settings (Table~\ref{table:nota_cross_ns}, improvement of 46\% on average when using \mcdropout{}) \textbf{which answers our last research questions that the uncertainty estimates from stochastic neural rankers do improve the effectiveness of the NOTA prediction task (by an average of 33\% for all conditions considered).}
\section{Conclusions}
In this work we study the calibration and uncertainty estimation of neural rankers, specifically \bertRankers{}. We first show that deterministic \bertRanker{} is not robustly calibrated for the task of conversation response ranking and we improve its calibration with two techniques to estimate uncertainty through \emph{stochastic neural ranking}. We also show the benefits of estimating uncertainty using risk-aware neural ranking and for predicting unanswerable conversational contexts. As future work, investigating other applications of stochastic neural rankers are important, e.g.  for other neural \ltr{} architectures, for other retrieval tasks~\cite{guo2019deep}, for fair retrieval~\cite{fern2020evaluating}, for ensembling neural rankers~\cite{zuccon2011back} and for query reformulation~\cite{lin2020query}.

\section*{Acknowledgements}
This research has been supported by NWO projects SearchX (639.022.722) and NWO Aspasia (015.013.027).


\bibliography{paper}
\bibliographystyle{acl_natbib}

\end{document}